# DARK MATTER AT VISCOUS-GRAVITATIONAL SCHWARZ SCALES: THEORY AND OBSERVATIONS

CARL H. GIBSON

University of California at San Diego, Departments of Applied Mechanics
and Engineering Sciences and Scripps Institution of Oceanography,
La Jolla, CA 92093-0411, USA, cgibson@ucsd.edu

The Jeans criterion for the minimum self-gravitational condensation scale is extended to include the possibility of condensation on non-acoustic density nuclei at Schwarz scales, where structure formation begins in the plasma epoch at proto-supercluster masses about 10,000 years after the Big Bang, decreasing to galaxy masses at 300,000 years. Then the plasma universe became relatively inviscid gas and condensed to $10^{23-26}$ kg "primordial fog particle" (PFP) masses. Baryonic dark matter by this theory should be mostly non-aggregated PFPs that persist in galactic halos. Schild (1996) suggests from quasar Q0957+561 microlensing that "rogue planets" are "likely to be the missing mass" of the lens galaxy. Non-baryonic dark matter composed of weakly interacting massive particles (WIMPs) should condense slowly at large viscous Schwarz scales to form galaxy supercluster halos, and massive galaxy cluster halos as observed by Tyson and Fischer (1995) for the rich galaxy cluster Abel 1689.

## 1 Introduction

Isaac Newton[1] suggested that "... if the matter [of the universe] were evenly disposed throughout an infinite space, it could never convene into one mass; but some of it would convene into one mass and some into another, so as to make an infinite number of great masses, scattered great distances from one to another throughout all that infinite space." Jeans[2,3] estimated the size of these "great masses" by assuming a pair of density perturbations with compensating pressures, propagating in a gas of density with the speed of sound $V_s$. By means of linear perturbation analysis he found that the amplitude of perturbations separated by distances greater than $L_J = V_s/(\ G)^{1/2}$ would grow and those shorter would not, where G is Newton's gravitational constant $6.7 \times 10^{-11}$ m$^3$ kg$^{-1}$ s$^{-2}$. Sound waves shorter than the Jeans length propagate more than a wavelength in the gravitational free-fall time $(\ G)^{-1/2}$ so there is no time for them to grow. Those much longer grow in density amplitude without limit. Jeans[3] [p. 345] claimed his criterion "provides simple, adequate and natural explanations for the great nebulae, the stars, the planets and the satellites of the planets." Cosmology studies[4,5,6,7,8,9,10] rely exclusively on Jeans criterion, although it is known to overestimate the rate of star formation in molecular clouds of the Milky Way spiral arms by about a hundred because of turbulence effects[11]. Here it is suggested that Jeans errors for cosmology are generally much larger and opposite in sign to Jeans errors for star formation. For the early universe, use of the Jeans mass criterion overestimates the minimum masses required for structure formation by factors of thousands to trillions.

Jeans' theory fails to consider nonacoustic density perturbations. One need not assume, as Jeans did, that all density perturbations propagate with the speed of sound. Gravitational condensation on a nonacoustic density nucleus in an infinite stagnant fluid of constant density and pressure is absolutely unstable. All the matter surrounding such a density nucleus is attracted towards it and will start moving in its direction. If mechanisms exist that permit the matter to stick to the nucleus in time periods t' shorter than the gravitational collapse time $\tau = (\rho G)^{-1/2}$ then its mass will grow indefinitely. Figure 1 compares the evolution of sinusoidal nonacoustic and acoustic density fluctuations, with wavelengths $\lambda$ smaller and larger than $L_J$, and with $t' < \tau$.

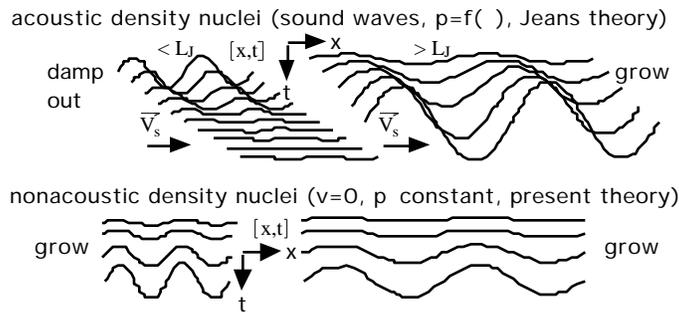

Figure 1: Nonacoustic and acoustic density fluctuations with wavelengths $\lambda$ smaller and larger than the Jeans length $L_J$. Nonacoustic waves grow for all $\lambda$.

As shown in Fig. 1, acoustic nuclei in a stagnant fluid grow only if their separations are larger than $L_J$. Nonacoustic density nuclei grow at all sizes. If the fluid is in motion, then viscous, turbulent (inertial) or other forces may arise to prevent the condensation. Gibson[12] considers effects of viscous and turbulent forces on self-gravitational condensation, and the cosmological consequences[13,14]. Figure 2 illustrates condensation on density nuclei limited by either viscous or turbulent forces. Constant pressure p nonacoustic density nuclei of scale L are shown with twice the ambient density $\rho$. Viscous forces $F_V$ on the left balance gravitational forces $F_G$ at the viscous Schwarz scale $F_V = (\nu\gamma/\rho G)^{1/2}$, where $\nu$ is the kinematic viscosity of the fluid, and $\gamma$ is the rate-of-strain—shown as a uniform expansion in Fig. 2. For strong turbulence, inertial forces $F_I$ may be stronger than the viscous forces at the scale of the gravitational condensation, so the scale will be the turbulent Schwarz scale $L_{ST} = (\varepsilon)^{1/2}/(\rho G)^{3/4}$ as shown on the right using Kolmogorov's second hypothesis, where $\varepsilon$ is the viscous dissipation rate of the turbulence. A transition length scale $L_{GIV} = (\nu^2/\rho G)^{1/4}$ exists, where gravitational, inertial, and viscous forces are all equal at the condensation length scale.

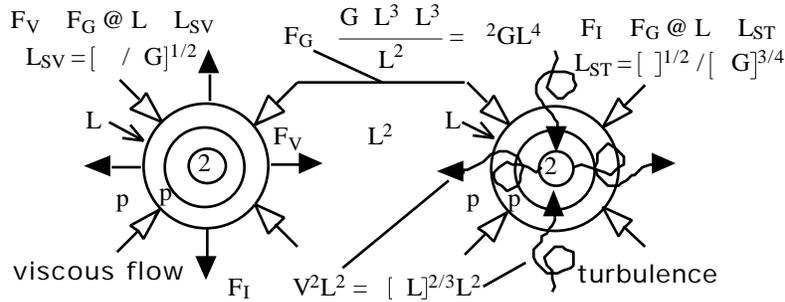

Figure 2: Condensation length scales $L_{SV}$ and $L_{ST}$ for nonacoustic density nuclei, dominated by viscous forces on the left and turbulent forces on the right.

Structure formation using the Jeans criterion and baryonic matter alone cannot account for the universe as we see it. No condensation is possible by the Jeans criterion in the plasma epoch before 300,000 years after the Big Bang because the speed of sound $V_s$ is $\sim 1/\sqrt{3}$ the speed of light c, so the Jeans mass $M_J = L_J^3 \rho$ is larger than the Hubble mass $c^3 t^3 \rho$ during this period, where t is the age of the universe. Measurements of the Cosmic Background Radiation (CBR) shows normalized temperature, and therefore density and velocity, fluctuations were on the order of $10^{-5}$ or less. Densities had decreased to such low levels that condensation times $\tau$ were larger than t. Even after decoupling the Jeans mass remains enormous, about $10^{36}$ kg[5], the mass of a globular cluster of stars. To obtain baryonic structures, decoupled cold dark matter, warm dark matter, and mixed dark matter structures are postulated on which the baryonic matter might condense in time to form the structures observed. Padmanabhan[8] assumes the first stars formed on such nonbaryonic nuclei, which are assumed to begin their collapse at about $3 \times 10^{11}$ s when the mass of the universe begins to dominate energy.

Unfortunately, the physical properties that must be postulated for the "weakly interacting massive particles" (WIMPs) of nonbaryonic dark matter are inconsistent with such an early, rapid condensation to form small scale nuclei. Both the "sticking times" t' and the kinematic viscosities $\nu$ for WIMP fluids should be much larger than those for baryonic fluids, considering the small collision cross sections ($<10^{-40}$ m$^2$) required to explain their low level of detectability. Consequently, WIMP gravitational condensation times should be slower and condensation lengths should be larger than those for baryonic condensation. By the present theory, the nucleation roles of baryonic and nonbaryonic matter are reversed compared to those generally assumed[8], with nonbaryonic matter gradually forming halos on the largest baryonic structures as nuclei.

## 2  Structure Formation in the Universe

The largest mass objects in the universe are superclusters of galaxies, which have a foam-like topology with "great walls" enclosing "supervoids".  The scale of the largest supervoid[9] is about $10^{24}$ m, much smaller than the present horizon scale of over $10^{26}$ m. This maximum size object in the universe also corresponds to the peak of the CBR density fluctuation power spectra observed by COBE, and the same peak derived from galaxy redshift catalogs.  Therefore, the indicated mass of the largest supercluster $M_{SC}$, assuming a flat universe with present density of $10^{-26}$ kg m$^{-3}$ is about $10^{47}$ kg, or a millionth of the critical mass of $10^{53}$ kg within our present horizon.  We take $M_{SC}$ to be a fossil of the first self-gravitational condensation event that occurred as the viscous Schwarz scale $L_{SV}$ decreased to equal the increasing Hubble scale $L_H = ct$.  We estimate the Hubble mass $\rho(t)c^3 t^3$ from $\rho(t) = \rho_o/a(t)^3$ computed from the Einstein field equations[15] and presented by Weinberg[4] as the standard cosmological model, as shown in Figure 2, where o-subscripts denote the present time and $a(t)$ is the scale factor $R(t)/R_o$ for any length R.

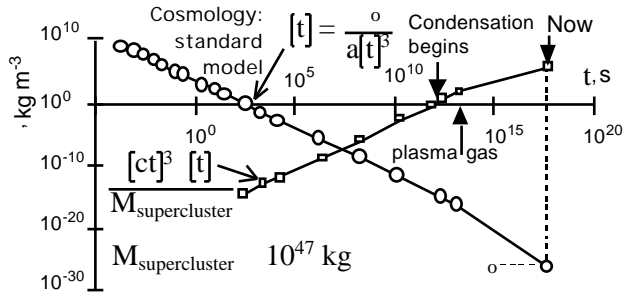

Figure 3: The time of first condensation is $10^{12}$ s by equating the supercluster mass $M_{SC} = 10^{47}$ kg to the Hubble mass from the cosmology standard model[4].

As shown in Fig. 3, the ratio $\rho(t)c^3 t^3/M_{SC}$ crosses 1 at time $10^{12}$ s, or 30,000 y, indicating that this is the time when the first condensation began.  This is earlier by a factor of ten than usually assumed for baryonic matter, with first condensation mass larger by a factor of $10^{11}$.  The kinematic viscosity $\nu$ of the fluid indicated by equating $L_H = L_{SV}$ at $10^{12}$ s is about $10^{28}$ m$^2$s$^{-1}$, giving a subcritical Reynolds number $L_H^2/\nu t = (3\times 10^{20})^2 \, 10^{-12}/10^{28} = 9$ at the condensation scale, consistent with the assumptions that the flow is limited by viscous forces at $L_H$, with $\gamma = 1/t$ that of the expanding universe. Since we see that these supercluster mass "condensates" have expanded from their horizon scales of about $3\times 10^{20}$ m at $t=10^{12}$ s to their present sizes of $10^{24}$ m, at average velocity 0.01c, the effect of gravity on these structures has been to cause a deceleration of their expansion rate. The viscosity of $= 10^{28}$ m$^2$s$^{-1}$ is slightly smaller than $c^2 t = 9 \times 10^{28}$ m$^2$s$^{-1}$, which is the maximum physically meaningful value within the horizon.

As the plasma expands it cools, and the ratio $L_{SV}/L_H$ decreases from its value of 1 at $t=10^{12}$ s to values about $10^{-5}$ at the time of plasma neutralization at $t=10^{13}$ s, corresponding to galaxy masses of $10^{42}$ kg. The horizon scale was $3 \times 10^{21}$ m for these proto-galaxy condensates. The viscous Schwarz scale was $L_{SV} = (M_{gal}/\ )^{1/3} = 4.4 \times 10^{19}$ m, giving a viscosity $= 1.3 \times 10^{25}$ m$^2$s$^{-1}$. The corresponding Reynolds number $L_{SV}^2 / = 15$, where $= 10^{-13}$ s$^{-1}$, which is still subcritical, confirming our assumption that the flow is nonturbulent at the condensation scale. The size of galaxies now is similar to the size of the horizon then, or about $10^{21}$ m, so these have expanded their scale at an average velocity of $2 \times 10^{-5}$ c. Thus galaxies have continued to expand their sizes, but much less rapidly than clusters or superclusters since their gravitational deceleration began.

After plasma neutralization to gas, the viscosity decreased remarkably. The viscosity of a primordial mixture of hydrogen and helium gas at 20°C and atmospheric pressure is $\mu_{293} = 10^{-5}$ kg m$^{-1}$ s$^{-1}$, with temperature dependence $\mu/\mu_{293} = (T/293)^{0.67}$ up to T = 1000 K. Extrapolating to 3000 K gives $\mu = 4.75 \times 10^{-5}$ kg m$^{-1}$ s$^{-1}$, and $5 \times 10^{12}$ m$^2$ s$^{-1}$ with $10^{-17}$ kg m$^{-3}$. Assuming continued laminar flow, $L_{SV} = (\ /G)^{1/2} = (5 \times 10^{12} \times 10^{-13}/10^{-17} \times 6.7 \times 10^{-11})^{1/2} = 2.7 \times 10^{13}$ m, so the condensation mass is $L_{SV}^3\ = (2.7 \times 10^{13})^3 \times 10^{-17} = 2 \times 10^{23}$ kg, which is smaller than the Jeans mass by a factor of about $10^{13}$ and comparable to the mass of the earth's moon ($7.4 \times 10^{22}$ kg). Figure 4 shows a schematic of the proposed structure formation scenario.

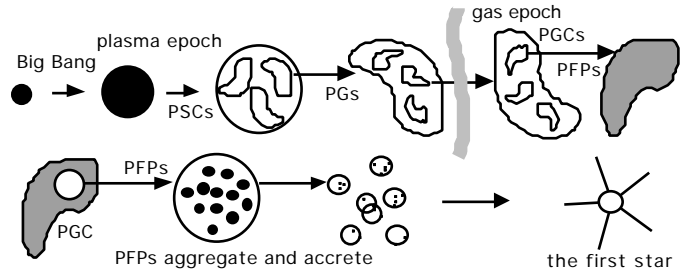

Figure 4: Structure evolution scenario with gravitational condensation on nonacoustic density nuclei at Schwarz scale $L_{SV}$, as described in the text.

The "rogue moon" objects with largest   began their formation promptly after plasma neutralization, and have been compacting and cooling ever since. Because they were so small compared to all other structures, and because they formed from primordial gas, it seems appropriate to refer to them as "primordial fog particles" (PFPs). The entire universe at $10^{13}$ s (300,000 y) turned to fog, in the beginning, filling in with 100% PFP dark matter the proto-globular-cluster (PGC), galaxy (PG), cluster (PC) and supercluster (PSC) gas structures that began deceleration during the plasma epoch. These PFPs are the basic material of construction for all subsequent larger structures. Detailed dynamical mechanisms by which they compacted, clustered, and accreted to form the first stars is an important topic for further study. PGC droplets have $_{max}$, from which the first PFPs should form in minimum time $= (G)^{-1/2}$, and with the smallest possible masses $M_{PFP} = (\mu /G)^{3/2}/ ^2$. Because PFPs become increasingly collisionless as they compact, it follows that the smallest particles will aggregate most rapidly, so the first stars to form should appear within the PGC droplets with maximum density. It also follows that the remaining regions of the universe with lower density should form larger mass PFPs, separated by larger distances. Star formation in such regions should be slower than in PGCs, and since we have seen that the PG, PC, and PSC structures continue to expand with the universe, it seem likely that most such PFPs that fail to form stars in the early moments of the universe, when they were sufficiently collisional to accrete and cluster with others, will remain permanently as baryonic dark matter. It also follows that if any PG gas droplets had smaller than any others, their PFP condensation would be delayed, their PFP masses would be large and widely separated, and the entire galaxy would likely remain dim or dark, with few or no stars, and many PFP "rogue Jupiters".

## 3  Comparison to Observations

The condensation scenario of Fig. 4 is quite different from scenarios based entirely on the Jeans criterion. By the present model, structure formation begins during the plasma epoch, but is forbidden by the Jeans criterion. Proto- superclusters, clusters and galaxies pre-formed in the plasma epoch are filled from the bottom up during the gas epoch, starting with primordial fog particles. These should rapidly aggregate to form the first small, long lived, stars in the most dense proto globular star clusters. Such small stars can form because the previous formation process was incredibly gentle, with little or no turbulence. Conventional cosmology based on Jeans instability criterion requires a turbulent collapse with fragmentation of turbulent Jeans mass gas clouds[10]. No small stable stars can form by the Jeans scenario because $L_{ST}$ values would be enormous.

       Globular clusters within protogalaxies continue to aggregate and accrete by the same weakly collisional, stellar dynamical mechanisms guiding the aggregation and accretion of the PFP clusters from which they were formed. Similarly, massive elliptical galaxies form by aggregation of smaller elliptical galaxies. Because the process is a self similar nonlinear cascade from small scales to large, the distribution function of density should be lognormal with intermittency factor (variance of ln ) increasing with time as the range of scales of the condensed structures increases. Extreme intermittency of averaged over small scales can have important consequences in sampling PFP dark matter,

since it will no longer be uniformly distributed in regions of a galaxy where a wide range of structure masses have formed. For this reason, confidence intervals on exclusion diagrams derived from star microlensing in the MACHO and EROS collaborations, assuming a uniform dispersion of MACHO particles in various mass ranges, should be interpreted with caution. No information is apparently available about the intermittency of the spatial distribution of MACHO particles.

Quasar microlensing of galaxy halos have the advantage of averaging over larger scales for which clumping effects are less important. Optical densities; that is, probabilities of detecting microlensing objects of a particular size along a single line of sight, are typically a million times the optical density for star microlensing. Robust microlensing events are more probable using the small, intense, light sources of quasars to search for microlensing events in the halos of lens galaxies thousands or millions of times more distant, than for similar objects in our own galaxy halo. Although preliminary MACHO and EROS exclusion diagrams suggest MACHOs down to $10^{23}$ kg cannot provide more than 20% of the halo mass, Schild[16] reports continuous microlensing by objects of $10^{25}$ kg mass that are "likely to be the missing mass" based on nearly 1000 nights of observations of Q0957+561 A,B triple image brightness pairs of the double image lensed quasar. Refsdal and Stabell[17] reach the same conclusion for Q2237+0305, but suggest more accurate light curves over longer time spans be provided, such as given by Schild[16]. Hawkins[18] examines several lensed quasars, and notes that most are lensed by dark galaxies with mass/luminosity ratios of order 1000, with microlensing masses usually near $10^{26}$ kg or more that can account for all of the galactic dark matter.

O'Dell and Handron[19] show Hubble Space Telescope photographs of "cometary globules" in the Helix planetary nebula that may actually be PFPs that have been made visible by the powerful gaseous shells emitted by the central dying star, which cause the volatile H-He liquid or frozen gases to evaporate and form a cometary envelop. The objects have masses of $10^{25}$ kg and are separated by about $10^{14}$ kg, as expected for primordial fog particles. Other planetary nebulae have shown evidence of similar objects, but are typically much more distant.

The first calibrated mass profile of a galaxy cluster[20] gives a smooth mass profile, $10^{45}$ kg, M/L = 400/1,  = $3 \times 10^{-21}$ kg m$^{-3}$, and $_{WIMP}$ = $10^{30}$ m$^2$ s$^{-1}$ setting $L_{SV}$ equal to the cluster core radius $6 \times 10^{21}$ m with  = $10^{-17}$ s$^{-1}$ (see Fig. 2).

## 4 Summary

A hydrodynamic analysis of the self gravitational structure formation process indicates that the classical Jeans instability criterion is incomplete because nonacoustic density nuclei have been neglected. Condensation on such nonpropagating nuclei is absolutely unstable, and is limited at viscous and turbulent Schwarz scales. By these criteria, the first condensations occur in the plasma epoch at supercluster to galaxy masses, followed by a condensation to form "primordial fog particles" proposed as the primary constituent of baryonic dark matter. Most PFPs should persist as "rogue moon-planet" dark matter in halos of galaxies, dominating the halo mass. Quasar microlensing observations of Schild[16] and others[17,18] reveal such halos. Nonbaryonic dark matter WIMP fluids are superviscous and lack the sticking mechanisms of baryonic matter, so they should have large $L_{SV}$ scales, and gradually condense on the largest baryonic structures as supercluster "superhalos" or "cluster halos". Cluster lensing[20] supports these model predictions.